\documentclass[english,aps,prl,twocolumn]{revtex4}
\usepackage[T1]{fontenc}
\usepackage[latin9]{inputenc}
\usepackage{amsmath}
\usepackage{amssymb}
\usepackage{graphicx}
\usepackage{epstopdf}

\makeatletter
\@ifundefined{textcolor}{}
{%
 \definecolor{BLACK}{gray}{0}
 \definecolor{WHITE}{gray}{1}
 \definecolor{RED}{rgb}{1,0,0}
 \definecolor{GREEN}{rgb}{0,1,0}
 \definecolor{BLUE}{rgb}{0,0,1}
 \definecolor{CYAN}{cmyk}{1,0,0,0}
 \definecolor{MAGENTA}{cmyk}{0,1,0,0}
 \definecolor{YELLOW}{cmyk}{0,0,1,0}
}

\makeatother

\usepackage{babel}
\begin{document}

\title{ Fermi-edge transmission resonance in graphene driven by a single Coulomb impurity}

\author{Paritosh Karnatak\footnotemark[1], Srijit Goswami\footnote{Equal Contributions. \newline email: paritosh@physics.iisc.ernet.in, s.goswami@tudelft.nl}\footnote{Current address: Kavli Institute of Nanoscience, Delft University of Technology, P.O. Box 5046, 2600 GA Delft, The Netherlands.}, Vidya Kochat, Atindra Nath Pal\footnote{Current address: Solid State Physics Laboratory, ETH Z\"{u}rich, 8093 Z\"{u}rich, Switzerland.}, Arindam Ghosh}

\affiliation{Department of Physics, Indian Institute of Science, Bangalore 560012, India}

\begin{abstract}
The interaction between the Fermi sea of conduction electrons and a non-adiabatic attractive impurity potential can lead to a power-law
divergence in the tunneling probability of charge through the impurity. The resulting effect, known as the Fermi edge singularity (FES), constitutes one of the most fundamental many-body phenomena in quantum solid state physics. Here we report the first observation of FES for Dirac Fermions in graphene driven by isolated Coulomb impurities in the conduction channel. In high-mobility graphene devices on hexagonal boron nitride substrates, the FES manifests in abrupt changes in conductance with a large magnitude $\approx e^{2}/h$ at resonance, indicating total many-body screening of a local Coulomb impurity with fluctuating charge occupancy. Furthermore, we exploit the extreme sensitivity of graphene to individual Coulomb impurities, and demonstrate a new defect-spectroscopy tool to investigate strongly correlated phases in graphene in the quantum Hall regime.
\end{abstract}

\maketitle
\selectlanguage{english}%

Experimental observation of interaction-driven many-body phenomena
in graphene depends critically on background disorder, which in mechanically
exfoliated graphene arises primarily from the trapped Coulomb impurities
at the substrate or at the surface (adatoms or adsorbates)~\cite{Graphene_neto_RMP,K_adsorbates_nphys}.
The discovery of the fractional quantum Hall effect~\cite{fqhe_EvaAndrei,fqhe_Kim}, broken symmetry states at zero filling factor~\cite{Yacoby_doublygatedsus_bilayer},
or spin-valley quantum Hall ferromagnetism~\cite{Kim_spinvalley},
have all been achieved in high mobility suspended graphene devices or graphene
on hexagonal boron nitride (hBN) substrates, where the influence of
external disorder was minimized. However, the ability of graphene
to screen an external impurity potential, not only determines the
strength of electron-electron interaction and related physical properties
such as the dielectric constant~\cite{STM_crommie} or the Fermi
velocity~\cite{DCreshapeGeim}, it also induces another class of
many body effects such as the Anderson orthogonality catastrophe~\cite{OrthoCat_Anderson,OrthoCat_guinea},
local moment formation~\cite{LocMagMom_castro,loc_spin_PRL} and
the Kondo effect~\cite{Kondo_Baskaran_prb,Kondo_Fuhrer,kondo_landau}. This latter
class of phenomena, where disorder itself plays a central role, remains
poorly explored in graphene.

Some experimental attempts have now been made to probe many-body
effects induced by disorder in graphene. These include scanning tunneling
spectroscopy of isolated Coulomb impurities adsorbed on graphene~\cite{STM_crommie,STM_crommie2,Eva_CI_PRL},
and quantum transport via localized states created by ion irradiation~\cite{Kondo_Fuhrer}.
However, the structural integrity of graphene in the latter can be
affected too adversely to identify many-body effects unambiguously.
In this letter we report a striking new phenomenon in the electrical
transport in high-mobility graphene, on hBN, which manifests in random
telegraph switching in the electrical conductance with a large
magnitude of $\approx e^{2}/h$ at resonance. The switching
is observable only within restricted windows of gate voltage, and
the analysis of the switching rates suggests tunneling of charge,
driven by attractive Coulomb interaction between the Dirac Fermions
in graphene and unintentional Coulomb impurities located near one
of the metallic contacts. We suggest that this remarkable effect represents
the first evidence of a many-body transmission resonance via Fermi-edge
singularity in graphene. Although FES is immensely
important to many phenomena including X-ray absorption~\cite{xray_RMP},
Kondo effect~\cite{kondo_Anderson_PRB,kondo_Andrei_RMP} and transport
in mesoscopic systems~\cite{Cobden_FES_prl,FES_Geim}, it has thus far never
been observed in graphene, where the Dirac-fermionic charge carriers
may play a nontrivial role~\cite{OrthoCat_guinea}.

\begin{figure}[th]
\begin{centering}
\includegraphics[width=1\linewidth]{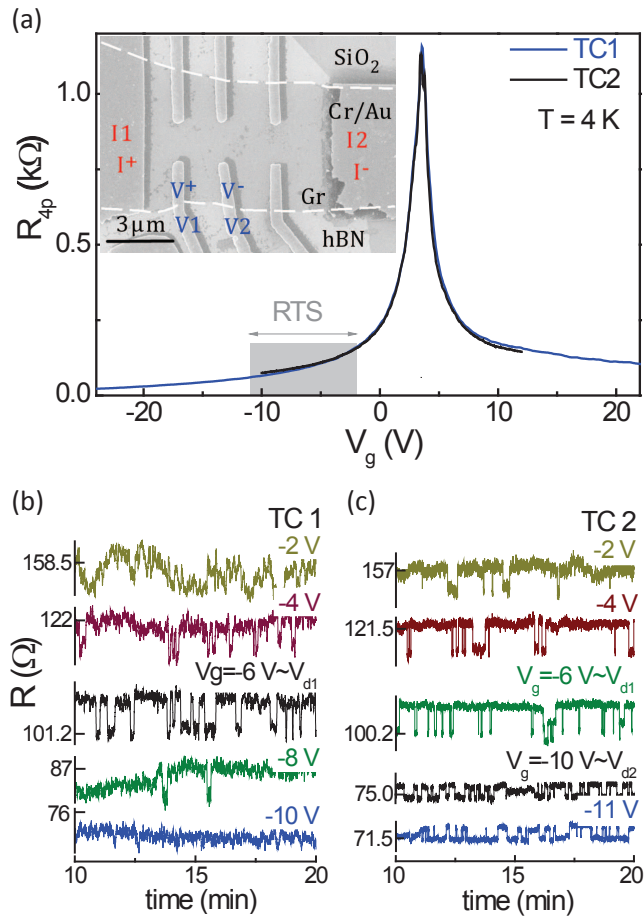}
\par\end{centering}

\caption{Random telegraph signal (RTS) in graphene on hexagonal boron nitride
(hBN). (a) Resistance ($R$) vs. back gate voltage ($V_{g}$) characteristics
of the device at $4$~K for two thermal cycles (TC1 and TC2). Shaded
region shows the $V_{g}$ range in which RTS is observed. Inset shows
the electron microscope image of the device and electrical leads.
(b) Resistance time series for TC1 at different $V_{g}$ values show
an enhanced switching rate at $V_{g}\simeq-6$~V. (c) Resistance
time series displaying an enhanced switching rate at $V_{g}\simeq-10$~V
for TC2.}
\end{figure}

The graphene transistor explored in detail in this experiment was prepared on a
hBN substrate, closely following the approach in~\cite{Zomer_GBN}.
Details of the fabrication procedure can be found in the supplementary section.  Fig.~1a (inset) shows an electron microscope image of the device used in our experiments. Resistance ($R$) vs. back gate voltage ($V_{g}$) characteristics shown in Fig.~1a reveal electron and hole mobility to be $\approx 83,000$~cm$^{2}$/Vs and $\approx65,000$~cm$^{2}$/Vs, respectively. All measurements are carried out at $T = 4$~K, with a small source-drain bias, $V < k_{B}T/e$, such that the system is maintained in thermal equilibrium. The key result of this work is the unexpected observation of a random telegraph signal (RTS) in $R$ (current leads: I1-I2, voltage leads: V1-V2 in Fig.~1a) as a function of time, detectable only over a certain $V_{g}$ range (shaded region in Fig.~1a). For low $V_{g}$ close to the Dirac point, the RTS becomes hard to isolate from the background $1/f$ noise, while for $V_{g}\lesssim-11$~V the amplitude of the RTS was too small to detect. This explains why most earlier noise studies \cite{atin_acsnano_nmechanism,atin_cvd_apl,atin_MLG_ultralow_apl,atin_bilayer} in low or moderate mobility graphene failed to explore the RTS in detail, where large $1/f$ noise dominates resistivity noise over a wide range of gate voltage ($\gtrsim20$~V) on either side of the Dirac point.

In this experiment, we subject the device to several thermal cycles between $4$~K and $\sim 20$~K. This does not affect the $R-V_{g}$ characteristics significantly, but modifies the RTS characteristics. This is illustrated with two of the thermal cycles, denoted as TC1 and TC2 in Fig.~1, in which the RTS noise was most pronounced. The local enhancement of RTS amplitude and switching rate in both TC1 and TC2 (Fig.~1b,~c) indicates a resonant defect level around $V_{g}\sim-6$~V. An additional level appears at more negative $V_{g}$ for TC2 (Fig.~1c), causing detectable switching noise down to $V_{g}\sim-11$~V. The observation of RTS in high-mobility graphene transistors is an important result, as it implies that its electrical properties can be significantly modified even by a single charge trap or impurity.

\begin{figure*}[tbh]
\begin{centering}
\includegraphics[width=0.85\linewidth]{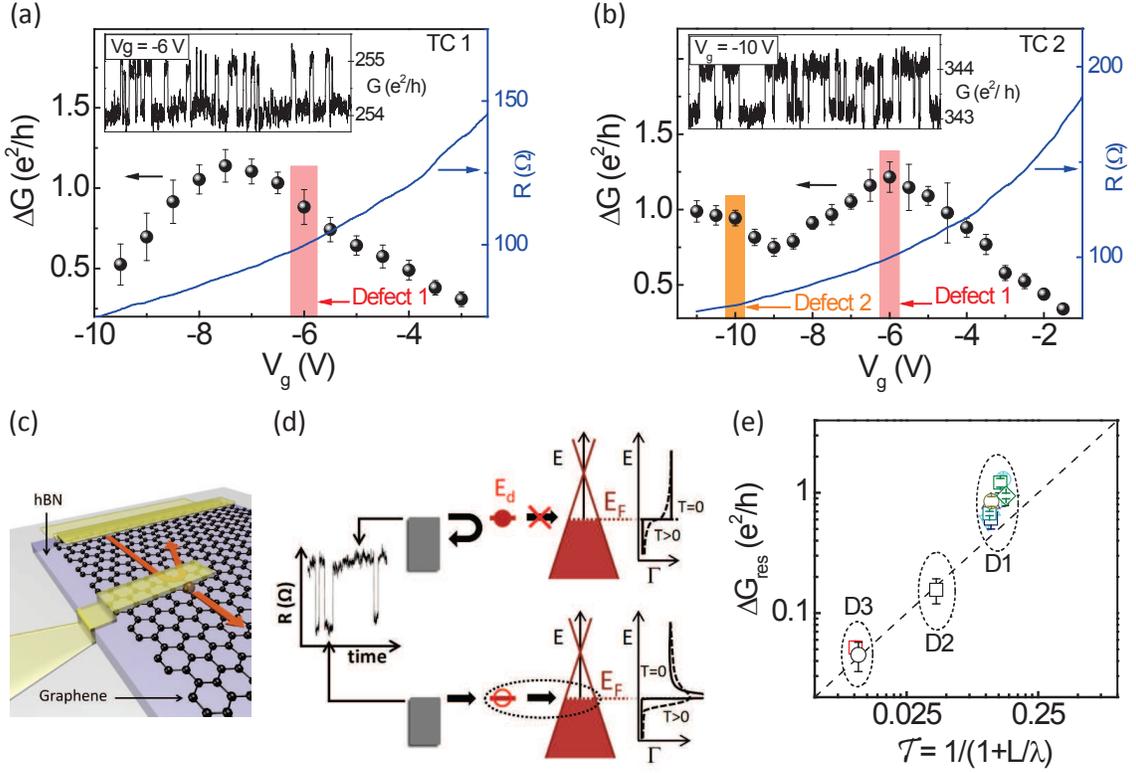}
\par\end{centering}

\caption{Magnitude and mechanism of RTS. (a) Magnitude of the conductance jumps ($\Delta G$) as a function of back gate voltage ($V_{g}$) for thermal cycle 1 (TC1). Solid line shows resistance as a function of $V_{g}$. Inset shows the time series conductance (in the units of $e^2/h$) at $V_{g} = -6$~V. (b) $\Delta G$ vs. $V_{g}$ for thermal cycle 2. Note that $\Delta G$ has two maxima of $\approx e^{2}/h$ at $V_{g}\simeq-10$~V and $V_{g}\simeq-6$~V, indicating the presence of two active defects. Inset shows conductance time series at $V_g = -10$~V. (c) An illustration of the device showing possible location of a charged defect close to one of the metal contacts. (d) Schematic of the microscopic mechanism, depicting the formation of a bound state between graphene and a Coulomb impurity. (e) The change in conductance at resonance for all samples as a function of the corresponding transmission probability ($\mathcal{T}$). The dashed line has a slope of $4$.}
\end{figure*}

RTS has often been observed due to carrier number fluctuation in nanoscale or mesoscopic semiconductors with small number of carriers~\cite{Noise_review_KirtonUren}, but this model fails here as the observed RTS amplitude ($\Delta R/R\sim0.1-0.5$~\%) is more than two orders of magnitude larger than the expected $\Delta R/R\sim10^{-5}$ for charge exchange with a single defect at (hole) doping of $\sim7\times10^{11}$~cm$^{-2}$ ($V_{g}=-6$~V). Mobility fluctuation due to a single two-level trap-state is not only a physically unlikely scenario, but it also yields $\Delta R/R < 10^{-4}$ assuming a reasonable Hooge parameter of $10^{-3}$~\cite{atin_acsnano_nmechanism,atin_bilayer,atin_cvd_apl,atin_MLG_ultralow_apl}. Finally, {\it no} RTS is observed when voltage leads other than V1 are used to measure $R$ (see supplementary material for details), allowing us to conclude that the observed RTS is not intrinsic to graphene and is observed only when the Fermi energy $E_{F}$ of graphene aligns with the discrete energy level of a defect or impurity that is located close to the lead V1 (see schematic in Fig.~2c), presumably incorporated unintentionally during device fabrication.

Experiments with small silicon MOSFETs~\cite{Cobden_FES_prl} show that strong RTS can also occur due to intermittent occupancy fluctuation of an impurity in proximity to the conducting 2D channel where the channel and the charged impurity  interact via an attractive Coulomb potential. When the impurity state is resonant with the Fermi energy of the channel, the interaction screens the charged impurity by forming a many-body bound state.  This causes a power-law singularity~\cite{Matveev_Larkin_prb}, called the Fermi edge singularity, in the tunneling probability between the impurity state and the 2D channel, in a manner that is analogous to the Kondo effect with localized spins. Effectively, the interaction adds one conduction channel via the impurity state which, in the unitary limit, increases conductance by as much as $ge^{2}/h$ (where $g$ is the level degeneracy) compared to when the impurity is neutral or repulsive.

To explore this possibility in our devices, we analyze the RTS amplitude as a function of $V_{g}$ for both thermal cycles. The time series in the insets of Figs.~2a and 2b demonstrate the RTS to correspond to change in conductance $\approx e^{2}/h$ at both resonances ($V_g \simeq -6$~V and $-10$~V). Figs.~2a and 2b show the average $\Delta G = |1/R_D -1/R_U|$ as a function of $V_{g}$, where $R_{U}$ and $R_{D}$ are the high and low resistance states of the RTS, respectively. $\Delta G$ varies nonmonotonically with $V_g$, and peaks at $V_{g}\simeq-6$~V (TC1 and TC2) and $V_{g}\simeq-10$~V (TC2), confirming the resonance-like behavior. Strikingly, the magnitude of the conductance change at resonance, $\Delta G_{res}$ is $\approx e^{2}/h$, and remains similar at all resonances, irrespective of impurity-specific details such as its location and coupling to graphene. Such a universality readily rules out a non-interacting mechanism of resonant tunnelling between the impurity state and graphene, but suggests a many-body effect with singular quasiparticle tunneling.

For a quantitative microscopic framework we assume screening of a Coulomb impurity located at the peripheral region of lead V1 (Fig.~2c). Intermittent tunneling out of a hole (\textit{i.e.} capturing of an electron) causes a fluctuating attractive Coulomb potential between the impurity and the 2D hole gas in graphene. As schematically described in Fig.~2d, this corresponds to a time-dependent two-state fluctuation in the \textquotedblleft{}contact\textquotedblright{} resistance, as the quasi-bound state of the charged defect and graphene incorporate one additional conduction channel. Including the scattering of carriers between the voltage leads, the net expected conductance change at resonance can be expressed as $\Delta G_{res} = (ge^2/h)\times \mathcal{T}$, where $\mathcal{T} \approx 1/(1 + L/\lambda)$, is the transmission probability of charge between the leads~\cite{Supriyo_El_transport}. Here $\lambda$ is the mean free path which can be readily estimated from the $R-V_{g}$ characteristics (see supplementary material for details), and $L$ is the separation between the leads.

\begin{figure}[tbh]
\begin{centering}
\includegraphics[width=0.9\linewidth]{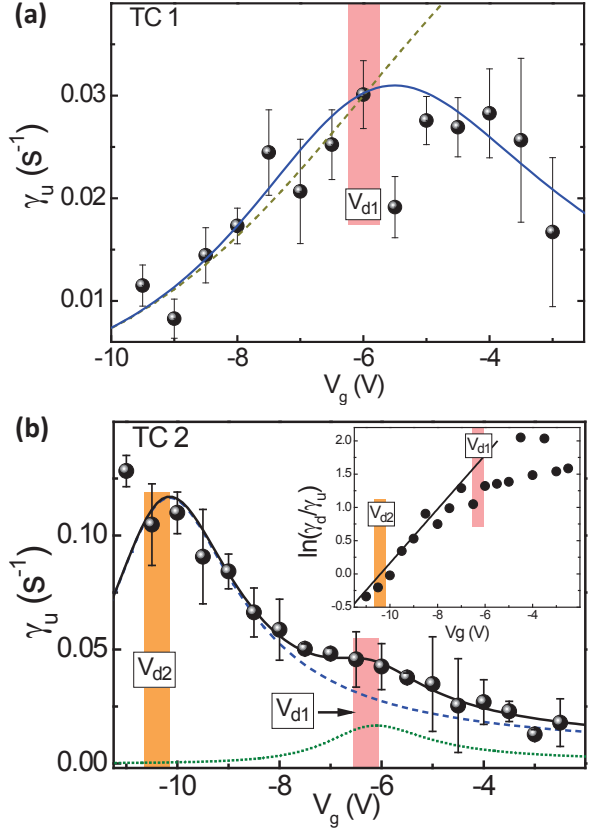}
\par\end{centering}

\caption{Gate voltage dependence of the transition rates. (a) The emission
rate (inverse of high resistance state lifetime) $\gamma_{U}$ vs.
$V_{g}$ in thermal cycle 1. The emission rate shows a peak near $V_{d1}$
indicating the presence of interaction effects. The solid line represents
Eq. (1), with $\alpha=0.2$, $\eta=0.00007$ and $V_{d1}=-6$~V and
the dashed line shows $\gamma_{U}$ expected from the non-interacting
model. (b) The emission rate $\gamma_{U}$ vs. $V_{g}$ in thermal
cycle 2. The emission rate peaks near $V_{d2}$ and $V_{d1}$, showing
contribution from both defects. The solid line shows the contribution
to the emission rate from both defects, it is a sum of Eq. (1) with
$\alpha=0.2$, $\eta=0.00014$ and $V_{d1}=-10.4$~V shown as the
blue dashed line and Eq. (1) with $\alpha=0.2$, $\eta=0.00018$ and
$V_{d1}=-6.3$~V shown as the green dotted line. Inset shows that
$\ln(\gamma_{d}/\gamma_{u})$ varies linearly with $V_{g}$, and the
slope changes near $V_{d1}$. Solid line yields the slope $\eta$,
used in the fitting above.}
\end{figure}

\begin{figure}[bh]
\begin{centering}
\includegraphics[width=1\linewidth]{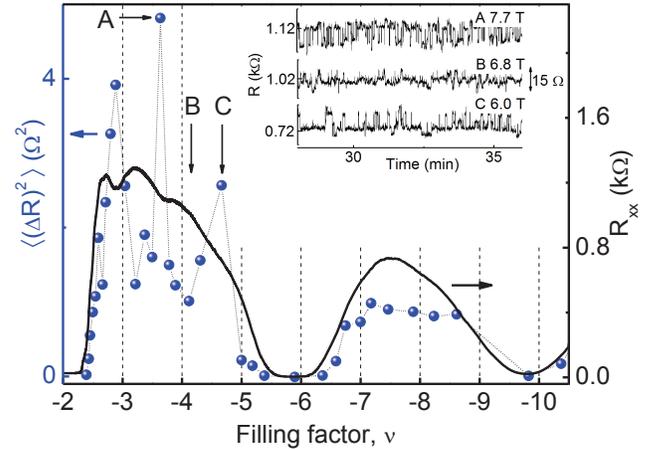}
\par\end{centering}

\caption{Defect spectroscopy of Landau levels (a) Variance of resistance (noise
magnitude), shown as solid spheres, closely follows the variation
in longitudinal resistance ($R_{xx}$, solid line). Strong
to moderate suppression in noise magnitude can be observed at most
integral filling factors (see text). The slightly shifted noise
minimum at $\nu = -3$ could be due to the discrepancy between local
filling factor and that estimated from the $B$-dependence of average
$R_{xx}$. 
Inset shows the nature of time series at different field values. }
\end{figure}

In Fig.~2e we have compiled the measured $\Delta G_{res}$ against the experimentally estimated $\mathcal{T}$ for all resonances observed in three different graphene devices. For the present high mobility device (D1), we observed 11 resonances spanning a $V_g$ range of $-1$~V to $-10$~V corresponding to $\mathcal{T} \sim 0.1 - 0.16$ ($\lambda \sim 200 - 350$~nm), over all the thermal cycles. In addition, we performed a similar analysis for two other devices, denoted as D2 and D3, where graphene was deposited directly on the SiO$_2$ substrate. The carrier mobility ($\sim 3000 - 4000$~cm$^2$/Vs) in these two devices is significantly lower, corresponding to $\lambda \sim 10 - 40$~nm, and they also displayed strongly suppressed $\Delta G_{res}$ (details of these devices and calculations can be found in the supplementary section). Most remarkably, as shown in Fig.~2e, the variation of the measured $\Delta G_{res}$ at all resonances in different devices falls close to $ge^2/h\times \mathcal{T}$ (dashed line) with $g = 4$, arising from the two-fold spin and valley degeneracy in graphene. This result emphasizes that RTS due to interacting Coulomb impurities could be generic to graphene field-effect devices which, counterintuitively, becomes stronger as the mobility of the channel is increased.

To further verify the role of interaction, we compute the emission
($\gamma_{U}$) and capture ($\gamma_{D}$) rate of charge (hole)
from the defect state as the inverse lifetimes of the $R_{U}$ and
$R_{D}$ resistance states, respectively (see Fig.~2d and supplementary material).
In Fig.~3a and 3b we have plotted the dependence of $\gamma_{U}$
on $V_{g}$ for thermal cycles TC1 and TC2, respectively. The $V_{g}$-dependence
of $\gamma_{D}$ for thermal cycles TC1 and TC2 is shown in the supplementary
material. The clear non-monotonic dependence of $\gamma_{U}$
on $V_{g}$ confirms the failure of the non-interacting framework.
In fact, we find that $\gamma_{U}$ follows the singular transmission
probabilities generalized for finite temperature (Fig.~3a,~b)~\cite{Cobden_FES_prl,Matveev_Larkin_prb},

\[
\negthickspace\gamma_{U,D}=AT^{\alpha-1}\exp\left(\pm\frac{E_{F}-E_{d}}{2k_{B}T}\right)\qquad
\]

\begin{equation}
\quad\qquad\qquad\qquad\times\frac{\left|\Gamma[\alpha/2+i(E_{F}-E_{d})/2\pi k_{B}T]\right|^{2}}{\Gamma(\alpha)}
\end{equation}

\noindent where $E_{F}-E_{d}\approx\eta e(V_{g}-V_{d})$ to the leading
order, with $E_{d}$ and $V_{d}$ being the defect energy and corresponding
gate voltage, respectively. The parameter $\eta$ couples the energy
difference to the offset in $V_{g}$, and is obtained directly from
the slope of $\ln(\gamma_{U}/\gamma_{D})$ as a function of $V_{g}$
in the vicinity of the resonances~\cite{Cobden_FES_prl}, as shown
in the inset of Fig.~3b. The analysis contains two fit parameters:
(1) The defect potential $V_{d}$, which was expectedly found close
to the $V_{g}$ corresponding to $\Delta G_{res}$ and $\gamma_{U}$ for each
resonance, and (2) the interaction parameter $\alpha$, which was
found to be $0.20\pm0.05$ for both resonances and thermal cycles.
The interaction parameter $\alpha$ governs the $T=0$ singularity
in transition rate $\sim\theta(E_{F}-E_{d})(E_{F}-E_{d})^{\alpha-1}$,
and is a direct measure of the Coulomb interaction between the defect
state and graphene ($\alpha=1$ represents the non-interacting case).
The observed $\alpha$ suggests a weak screening by graphene in spite of reasonably high doping ($\approx4-10\times10^{11}$~cm$^{-2}$)~\cite{OrthoCat_guinea}.
This is supported by the recent observation of low dielectric constant
of graphene even on hBN substrates~\cite{STM_crommie}, and affirms
the importance of including electron-electron interaction in evaluating
physical properties of graphene.

It is essential to comment on two key aspects of our experiment: First,
the Coulomb impurities in our device can arise from both structural
imperfections (such as edge defects or vacancies) or surface adsorbates,
in particular transition metal adatoms, such as Cr or Au,  perhaps 
incorporated during metallization.
The latter scenario is supported by (1) proximity of the active impurities
to one of the leads (voltage lead V1), and (2) occupation fluctuations
of the 3d atomic clusters (cobalt trimers) adsorbed on graphene have
been shown to occur in a similar range of gate voltage~\cite{STM_crommie}.
Secondly, resonances arising from bound state formation with singly charged impurities
via atomic collapse can also be important~\cite{OrthoCat_guinea,Wang_rydberg}
considering a low dielectric constant for graphene~\cite{STM_crommie}.
However, the RTS we observe does not exhibit features exclusive to the Dirac fermions. 

Subsequently, we employ the tunnel coupling between the Coulomb impurity
and graphene as a spectroscopic tool to explore the Landau level (LL)
structure in the integer quantum Hall regime, where Coulomb (exchange)
interaction may result in new broken symmetry states~\cite{QHF_macdonald_prl,Kim_spinvalley,Yacoby_doublygatedsus_bilayer,yacoby_sus_fqhe}.
The proximity of the impurity to the graphene channel makes defect
spectroscopy an extremely sensitive probe to the local density of
states (LDOS)~\cite{LL_spectroQD_Henini,LL_disordered_Luth}. In the experiment
here, we measured the RTS at different magnetic fields $B$ (perpendicular
to the plane of graphene) in the quantum Hall regime while keeping
the Fermi level fixed at $V_{g}=-6$~V (see typical traces in the
inset of Fig.~4). The RTS in the quantum Hall regime differ strongly
from the zero $B$ case and we find: (1) strong modulation in $\Delta R$
resulting from the wide variation in LDOS with $B$, (2) appearance
of multilevel structures, and (3) enhanced switching rate with multiple
characteristic rates. The overall variance of the resistance, $\langle(\Delta R)^{2}\rangle$,
was found to follow the Shubnikov-de Haas oscillations in longitudinal
resistance, $R_{xx}$, (Fig.~4) confirming that the RTS
is indeed sensitive to the LDOS, and that the RTS amplitude drops
sharply when the Fermi energy moves to a gap~\cite{Cobden_FES_prl}.

The striking observation in Fig.~4 is the clear suppression of the
RTS {[}minima in $\langle(\Delta R)^{2}\rangle${]} at filling factor
$\nu=-3$ and -4 in the first LL, and moderately at $\nu=-7$ and -8
in the second LL (the data at $\nu=-9$ was corrupted during the measurements),
suggesting removal of the four-fold spin-valley degeneracy even at
$B$ as low as $\sim 3$~T. The averaged resistance, $R_{xx}$ shows significantly weaker features at these fields establishing the superior sensitivity of single-defect noise spectroscopy in probing the LDOS. Recent transport experiments
at high magnetic fields attributed the lifting of degeneracy at half and quarter fillings to an
exchange-driven quantum Hall ferromagnetism in the SU(4) spin-valley
space~\cite{Kim_spinvalley}. The observation that the noise magnitude
indicates lifting of degeneracy, at these filling factors, 
even for Zeeman energy $E_{Z}\lesssim k_{B}T$, confirms strong orbital
effect of Coulomb interaction in forming the gaps. 

In conclusion, we have observed a new transmission resonance in high-mobility
graphene transistors, which manifests as a random telegraph signal
in conductance with amplitude as large as $ e^{2}/h$. Our
analysis suggests this to be due to Fermi edge singularity caused
by bound state formation between a local Coulomb impurity and charge
carriers in graphene. Apart from demonstrating a new many-body quantum
state in graphene, driven by individual Coulomb impurities, we have
also employed this phenomenon to probe the modification in the local
density of states of graphene by Coulomb interaction in the integer
quantum Hall regime.

\noindent We thank Andre Geim for useful discussions. S.G. thanks the
IISc Centenary Postdoctoral Fellowship for financial support. V.K. and A.N.P. thank CSIR for financial support. The authors thank the Department of Science and Technology for financial support.

%

\end{document}